\documentclass[aps,prd,groupedaddress,showpacs,nofootinbib]{revtex4}
\usepackage{graphicx,amssymb,amsmath,epsfig}
\usepackage[english]{babel}
\def\comment#1{}
\newcommand{\be}{\begin{equation}}\newcommand{\ee}{\end{equation}}
\newcommand{\bea}{\begin{eqnarray}}\newcommand{\eea}{\end{eqnarray}}
\newcommand{\beaa}{\begin{eqnarray}}\newcommand{\eeaa}{\end{eqnarray}}
\newcommand{\ba}{\begin{array}}\newcommand{\ea}{\end{array}}
\newcommand{\bit}{\begin{itemize}}\newcommand{\eit}{\end{itemize}}
\newcommand{\ben}{\begin{enumerate}}\newcommand{\een}{\end{enumerate}}

 \newcommand{\sfrac}[2]{\raisebox{0.095ex}{\scriptsize${\frac{#1}{#2}}$}}
 
 \newcommand{\sbf}[1]{\mbox{\scriptsize\bf{#1}}}

\begin{document}

\title{Melting of Wigner-like Lattice of Parallel Polarized Dipoles
}

\author{H. Kleinert
}


       \address{
Institut f\"ur Theoretische Physik,
 Freie Universit\"at Berlin,
Arnimallee 14, D14195 Berlin, Germany\\
ICRANeT, Piazzale della Republica 1, 10 -65122, Pescara, Italy
}


\vspace{2mm}

\begin{abstract}
We show that a
triangular lattice
consisting of
dipolar molecules
pointing
orthogonal to the plane
undergoes a first-order defect
melting transition.
\end{abstract}


\maketitle

About thirty years ago,
 Nelson
and
Halperin  \cite{HN}
extended the Kosterlitz-Thouless pair unbinding
theory
\cite{KO}
 of
vortices in a thin layer of superfluid helium
to
the phase transitions
of
defects
in two-dimensional
crystals.
They argued that
melting would
proceed by a sequence
of two Kosterlitz-Thouless transition,
the first when dislocations
of opposite Burgers vector unbind,
creating a hexatic phase,
and a second in which disclinations
of opposite Frank vector separate.
However, they never specified the physical
 parameter
of the crystal
which would decide
when this melting scenario
happens, rather than
a simple first-order melting transition
which was previously expected
on the basis of our three-dimensional experience.
Such a parameter
was found in
Ref.~\cite{SUCC},
and developed further
in \cite{SUCC2}, and
in the
textbook \cite{GFCM}.
It was shown that
a higher-gradient elastic constant called the
angular stiffness
determines which scenario takes place.
Only for a high angular stiffness
will the two-step melting process occur.
Otherwise
the melting transition would be
a completely normal first-order process.
Computer simulations
of the simplest lattice defect model on
a lattice confirmed the results \cite{JK}.

The theory was applied to a Lennard-Jones crystal and a
Wigner crystal,
and it was found
that in both cases the
angular stiffness was too small
to separate the melting transition into two successive
Kosterlitz-Thouless transitions.

Here we investigate the angular stiffness for
a crystal
that is  similar
to the Wigner crystal,
except that the repulsive forces
are due to parallel magnetic dipoles.
Thus the potential has the behavior $1/r^3$ rather than $1/r$.

\begin{figure}[tbhp]
\vspace{2cm}

\unitlength1mm
\def\fsz{\footnotesize}
\def\ssz{\scriptsize}
\def\tsz{\tiny}
\def\dst{\displaystyle}
\def\pu#1#2{\put(#1,#2){\emmoveto}}
\def\pd#1#2{\put(#1,#2){\emlineto}}
\begin{picture}(105.64,33.645)
\def\dst{\displaystyle}
\def\fsz{\footnotesize}
\def\IncludeEpsImg#1#2#3#4{\renewcommand{\epsfsize}[2]{#3##1}{\epsfbox{#4}}}
\put(0,0){\IncludeEpsImg{105.64mm}{26.43mm}{.250}{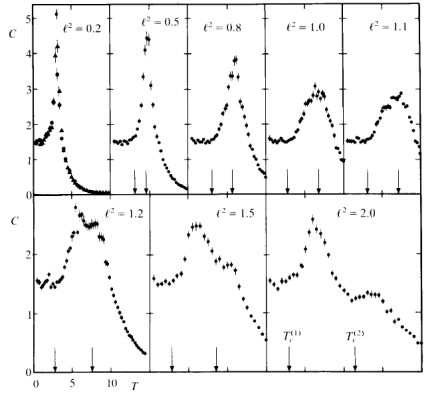}}
\end{picture}
\caption[]{Melting peaks in specific heat for
2 D lattice defect model showing the splitting of the melting transition
into two Kosterlitz-Thouless transitions if the parameter of
angular stiffness $\ell^2$ exceeds unity.
For small $\ell^2$, the transition is of first order,
for $\ell^2 \gtrsim  1$ it splits into two
Kosterlitz-Thouless  transitions.
The simulation data are from Ref.~\cite{JK}.
}
\label{@f}\end{figure}

The angular stiffness parameter
is defined as follows.
Let $\mu$ and $ \lambda $ be the usual elastic constants of
a crystal,
then the usual elastic energy
density depends on the
displacement field
 $u_i({\bf x})$ via the
 strain tensor
$u_{ij}({\bf x})\equiv
 [\partial _iu_j({\bf x})+
 \partial _ju_i({\bf x})]/2$
as follows
\begin{eqnarray}
 {\cal E} ={\mu}u_{ij}^2+\frac{ \lambda }{2}u_{jj}^2.
\label{@}\end{eqnarray}
The angular stiffness is parametrized
by the second of the higher-gradient energy
\begin{eqnarray}
  \Delta {\cal E} =
{ 2\mu }\ell^2 (\partial _i \omega _{j})^2
 +
\frac{ 2(\mu+\lambda) }{2}\ell'\hspace{1pt}^2 (\partial _iu_{jj})^2.
\label{@}\end{eqnarray}
where
$ \omega _j({\bf x})\equiv
\sfrac{1}{2} \epsilon _{jkl}\partial _ku_l({\bf x})$
is the local rotation field.
The parameter $\ell^2$ is the length scale of the angular stiffness.
It was argued in
\cite{SUCC} that for $\ell=0$,
dislocations
are indistinguishable from
neighboring pairs of
disclinations of opposite Frank vector,
and disclinations can be built from strings of dislocations.
There the transition
is of first order.
For high $\ell$, on the other hand,
beginning about with
the lattice spacing $a_0$,
the disclinations
could be suppressed
with the consequence that
the transition
based on disclination unbinding
would occur later than
that of dislocation unbinding.
The precise location of the
critical
$\ell$ was found by computer simulations \cite{JK},
and is plotted
in Fig.~\ref{@f}.

In order to apply this criterion
to the triangular lattice
formed by dipoles
we must calculate the elastic constants
in
 ${\cal E}$
and $ \Delta {\cal E}$.
This can be done for any repulsive interatomic potential
$\Phi({\bf x})=1/|{\bf x}|^p$ of power $p$, which
has the value $p=1$ for the Wigner crystal, and $p=3$
for the crystal of parallel dipoles.
For the second derivatives of this potential,
we calculate the Fourier transform
\begin{equation}
V_{ij}({\bf k})\equiv
\sum_{\sbf x\neq {\sbf 0}}[ 1-\cos( {\bf k}{\bf x})] \partial _i\partial _j \Phi({\bf x}).
\label{@VPOT}\end{equation}
\comment{
Alternatively we can use the Fourier transform of the potential
$\tilde \Phi({\bf k})$ and rewrite this as
as sum over reciprocal lattice vectors
\begin{equation}
V_{ij}({\bf k})\equiv
\sum_{\sbf c\neq {\sbf 0}}[
({\bf c}+{\bf k})_i
({\bf c}+{\bf k})_j \tilde\Phi(  {\bf c}+{\bf k})
-c_ic_j   \tilde\Phi(  {\bf c})
     ] .
\end{equation}
}
If $M$ is the mass of the
lattice constituents, the sound waves of polarization vector $
 \varepsilon_i ^{( \lambda)}({\bf k}) $ have the
frequencies determined by
\begin{eqnarray}
  \rho \omega ^{ (\lambda )}{}^2=V_{ij}
({\bf k})
 \varepsilon_i ^{( \lambda)}({\bf k})
 \varepsilon_j ^{( \lambda)}({\bf k}) ,
\label{@}\end{eqnarray}
where $ \rho $ is the mass density of the material.
This
can now be compared
with the equation of motion
following from the Lagrangian density
\begin{equation}
{\cal L}=\frac{ \rho }{2}\dot u_i^2({\bf x},t)
-
{   \cal E}
-{ \Delta    \cal E}.
\label{@}\end{equation}
which reads
\begin{eqnarray}
 \rho  \omega ^2 u_j^2({\bf k})-
\mu{\bf k}^2\left(1+\ell^2 {\bf k}^2\right)
P^T_{ij}({\bf k})
u_j({\bf k})
-(2\mu+ \lambda ){\bf k}^2
\left(1+\ell'\hspace{1pt}^2 {\bf k}^2\right)P^L_{ij}({\bf k})
u_j({\bf k})=0.
\label{@}\end{eqnarray}
where
\begin{equation}
P^T_{ij}({\bf k})
\equiv
\left( \delta _{ij}-\frac{k_ik_j}{{\bf k}^2}\right) ,~~~
P^L_{ij}({\bf k})
\equiv
\frac{k_ik_j}{{\bf k}^2} .
~~~
\label{@}\end{equation}
are the projections into transverse and longitudinal
directions with respect to ${\bf k}$.

Thus we merely have to calculate
the transverse part
of
(\ref{@VPOT})
and determine
$\ell^2$  from
 the
 ratio
of the ${\bf k}^2$-part versus the ${\bf k}^4$-part.
 This is not straight-forward.
Using the tensor decomposition
\begin{eqnarray}
\partial _i\partial _j\Phi({\bf x})=A \delta _{ij}+Bx_ix_j,
\label{@}\end{eqnarray}
with
\begin{equation}
A=\Phi'(r)/r=-p/r^{p+2},~~~B=\Phi''(r)/r^2-\Phi'(r)/r^3=p(p+2)r^{p+4},
\label{@}\end{equation}
we calculate
for small ${\bf k}$
\begin{eqnarray}
V({\bf k})= \sum_{\sbf x\neq {\sbf 0}}\left[
\frac{({\bf x}{\bf k})^2}2
-\frac{({\bf x}{\bf k})^4}{24}+\dots\right] (A \delta _{ij}+Bx_ix_j),
\label{@DEC}\end{eqnarray}
and find
\begin{eqnarray}
V_{ij}({\bf k})=
(
V_T^{(2)}k^2+
V_T^{(4)}k^4+\dots)( \delta _{ij}-
\hat{{\bf k}}_i
\hat{{\bf k}}_j)
+
(
V_L^{(2)}k^2+
V_L^{(4)}k^4+\dots)
\hat{{\bf k}}_i
\hat{{\bf k}}_j,
\end{eqnarray}
where
\begin{eqnarray}   \!\!\!\!\!\!\!\!\!\!\!\!\!\!\!\!
&&V_T^{(2)}=3A r^2/2+3Br^4/8=\frac{3p(2+3p)}{8}r^{-p},
~~~~~~~
V_T^{(4)}=-3A r^4/32-Br^6/128=\frac{p(10-p)}{128}r^{2-p},
\label{@}\end{eqnarray}
\begin{eqnarray}
&&V_L^{(2)}=3A r^2/2+9Br^4/8=\frac{3p(p-2)}{8}r^{-p},
~~~~~~~
V_L^{(4)}=-3A r^4/32-11Br^6/128=-\frac{p(10+11p)}{128}r^{2-p}.
\label{@}\end{eqnarray}
If the lattice sum would be carried out only over the
six nearest neighbors at $r=a_0$,
the length parameter of angular stiffness
would be given by
\begin{eqnarray}
{\ell ^2}\equiv
\frac{1}{{a_0^2}}\frac{
V_T^{(4)}
}{
V_T^{(2)}
}=\frac{10-p}{p-2}\frac{1}{48}.
\label{@}\end{eqnarray}
For dipole forces this is equal to
$\ell^2/a_0^2=7/48
\approx 0.145$.
Comparing this with
the phase diagram
of the general lattice defect model
shown in Fig.~\ref{@f},
we would conclude that
the melting transition is weakly of first order.

Let us now see the effect of the
full lattice sum.
Inspection of (\ref{@VPOT}) shows that the ${\bf k}^4$-part
cannot be calculated directly from the sum over
lattice sites,
since for $p=3$ the extra four powers
of ${\bf x}$ lead to a logarithmic
divergence.
To solve this problem
we set
$R=|{\bf x}+{\bf u}|$ and use Ewald's formula
to rewrite
\comment{\begin{eqnarray}
\sum_{\sbf x}
\frac{e^{i\sbf k\sbf x}}{R}=
\frac{1}{ \sqrt{\pi} }
\int _0^\infty
\frac{dt}{t} t^{1/2}
 \sum_{\sbf x}
{
e^{-t R^2+i\sbf k \sbf x}
}        ,
\label{@}\end{eqnarray}
or
\begin{eqnarray}
\sum_{\sbf x}
\frac{e^{i\sbf k\sbf x}}{R^3}=
\frac{2}{ \sqrt{\pi} }
\int _0^\infty
\frac{dt}{t} t^{3/2}
 \sum_{\sbf x}
{
e^{-t R^2+i\sbf k \sbf x}
}  ,
\label{@2}\end{eqnarray}
or more generally
}
\begin{eqnarray}
V_p({\bf k})=
\sum_{\sbf x}
\frac{e^{i\sbf k\sbf x}}{R^{^p}}=
\frac{1} {\Gamma (p/2) }
\int _0^\infty
\frac{dt}{t} t^{p/2}
 \sum_{\sbf x}
{
e^{-t R^2+i\sbf k \sbf x}
}
\,.
\label{@2}\end{eqnarray}
The sum over the lattice sites
 converges
fast  for large $t$.
For small $t$, it is convenient
to perform a duality transformation
that
 converts lattice sum into sums over the reciprocal lattice vectors:
${\bf c}$
\begin{eqnarray}
\sum_{\sbf x}
{e^{i\sbf k\sbf x}}= \frac{(2\pi)^D}{v}
\sum_{\sbf c}
 \delta ^{(D)}({\bf k}-{\bf c})\,.
\label{@RELA}\end{eqnarray}
Inserting this into the
Fourier
representation of an arbitrary function
\begin{eqnarray}
f({\bf x})=\int \frac{d^Dk}{(2\pi)^D} \tilde f({\bf k})e^{-i{\sbf x \sbf k}}\,,
\label{@}\end{eqnarray}
						       we obtain the relation
\begin{eqnarray}
 \sum_{\sbf x}  f({\bf x})
= \frac{(2\pi) ^D}{v}
\sum_{\sbf c}
\tilde f({\bf c}).
\label{@}\end{eqnarray}
The function
\begin{equation}
f({\bf x})=e^{-tR^2+i
{\sbf k}
{\sbf x}}
\label{@}\end{equation}
has a Fourier transform
\begin{equation}
\tilde f({\bf k}')=\frac{\pi^{D/2}}{t^{D/2}}
e^{-({\sbf k}'+{\sbf k})^2/4t+{\sbf k}' {\sbf u}}\,,
\label{@}\end{equation}
so that
\begin{eqnarray}
 \sum_{\sbf x}
e^{-tR^2+i
{\sbf k}
{\sbf x}}
 =
\frac{\pi^{D/2}}{vt^{D/2}}
\sum_{\sbf c}
e^{-({\sbf c}+{\sbf k})^2/4t+{\sbf c} {\sbf u}}\,.
\label{@}\end{eqnarray}
Inserting this into
Eq.~(\ref{@2})
we see that now the small-$t$ part
of the integrand converges fast.
For an optimal convergence
 we split the integrand
at some $t$-value $ \varepsilon $
and rewrite (\ref{@2})
as
\begin{eqnarray}
\sum_{\sbf x}
\frac{e^{i\sbf k\sbf x}}{R^{^p}}=
\frac{1}{ \Gamma (p/2) }\left[
\int _0^ \varepsilon
\frac{dt}{t} t^{p/2}
 \sum_{\sbf x}
{
e^{-t R^2+i\sbf k \sbf x}
}
+   \frac{\pi^D}{vt^{D/2}}
\int ^\infty_ \varepsilon
\frac{dt}{t} t^{p/2}
 \sum_{\sbf c}
{
e^{-({\sbf c}+{\sbf k})^2/4t+{\sbf c} {\sbf u}}
}                             \right] .
\label{@22}\end{eqnarray}
We now introduce the {\em Misra\/} functions
\begin{equation}
\varphi_n(z)\equiv
\int _1^\infty \,t^n e^{-zt}.
\label{@}\end{equation}
They are related to the incomplete Gamma functions
\begin{equation}
\Gamma( \alpha ,z)\equiv
\int _z^\infty \frac{dt}{t}t^{ \alpha }e^{-t}
\label{@}\end{equation}
by
\begin{equation}
\varphi_n(z)=z^{-n-1}\Gamma(n+1,z),
\label{@}\end{equation}
and can therefore be expanded as follows:
\begin{equation}
\varphi_n(z)=  z^{-n-1}\left[\Gamma(n+1)-\sum_{k=0}^\infty (-1)^k\frac{z^{k+n+1}}{k!(k+n+1)} \right].
\label{@xppa}\end{equation}
Using the
relation
\begin{equation}
\int _ \varepsilon ^\infty \,t^n e^{-zt}=
 \varepsilon ^{n+1}\varphi_n(z)\,,
\label{@xpa}\end{equation}
Eq. (\ref{@22})
can be written as
\begin{eqnarray}
V_p({\bf k})=
\sum_{\sbf x}
\frac{e^{i\sbf k\sbf x}}{R^{^p}}=
\frac{\varepsilon ^{p/2}
}{  \Gamma (p/2) }\left[ \varepsilon ^{p/2}
  \varphi_{p/2-1}
( \varepsilon R^2)e^{i{\sbf k}{\sbf x}}
+
   \frac{\pi^{D/2}}{v
 \varepsilon ^{D/2}}
 \sum_{\sbf c}
\varphi_{(D-p)/2-1}\left(
\frac{({\bf c}+{\bf k})^2}{4 \varepsilon }
\right) e^{i{\sbf k}{\sbf u} }
                             \right] .
 \label{@5}\end{eqnarray}
For a triangular lattice
with $D=2$,
lattice vectors
\begin{equation}
{\bf x}=a_0\left(l_1-\sfrac{1}{2} l_2,\sfrac{ \sqrt{3} }{2}l_2\right)
,~~~l_1,l_2={\rm all~integers},
\label{@}\end{equation}
cell volume $v= \sqrt{3}a_0^2/2$, and reciprocal lattice vectors
\begin{equation}
{\bf c}=\frac{2\pi}{a_0}\left(c_1,\sfrac{1}{ \sqrt{3} }c_1+
 \sqrt{ \sfrac{2}{ {3} }}c_2\right),
~~~c_1,c_2={\rm all~integers},
\label{@}\end{equation}
we choose $ \varepsilon =\pi/v$
so that the arguments
 $\varepsilon {\bf x}^2$ and
 ${\bf c}^2/4\varepsilon $   of the Misra functions
run through the same values
\begin{equation}
s\equiv  \epsilon {\bf x}^2
= \varepsilon a_0^2
\left(l_1-\sfrac{1}{2} l_2,\sfrac{ \sqrt{3} }{2}l_2\right) ^2
=\frac{2\pi}{ \sqrt{3} }(l_1^2-l_1l_2+l_2^2),\\
~ ~    s=\frac{{\bf c}^2}{4 \varepsilon }=\frac{(2\pi)^2}{a_0^2}
\frac{v}{4\pi}
\left(c_1,\sfrac{1}{ \sqrt{3} }c_1\!+\!
 \sqrt{ \sfrac{2}{ {3} }}c_2\right)^2\!\!= \frac{2\pi}{ \sqrt{3} }
(c_1^2+c_1c_2+c_2^2),
\label{@}\end{equation}
Then Eq.~(\ref{@5})
yields the formal relation
\begin{eqnarray}
\sum_{{\sbf x}}
\frac{1}{|{\bf x}|^p}=
\frac{\varepsilon ^{p/2}
}{  \Gamma (p/2) } \sum_{s}\left[   \varphi_{p/2-1}
( s)
+       \frac{\pi^{D/2}}{v
 \varepsilon ^{D/2}
}
\varphi_{(D-p)/2-1}\left(s
\right)
                             \right] .
 \label{@6}\end{eqnarray}
This becomes meaningful
by a subtraction of the \mbox{${\bf x}=0$~-term}, which yields
after a separate treatment of
the \mbox{$s=0$~-terms}
on the right-hand side, if we assume $p>D$,
\begin{eqnarray}
&&\!\!\!\!\!\!\!\!\!\!\!\!\!\!\sum_{\sbf x\neq {\sbf 0}}
\frac{1}{|{\bf x}|^p}= \lim _{{\sbf x}\rightarrow \sbf 0}
\left[-\frac{1}{|{\bf x}|^p}
+\frac{ \varepsilon ^{p/2}}{  \Gamma (p/2) } \sum_{s}   \varphi_{p/2-1}(
\varepsilon
 {\bf x}^2)\right]
+\frac{1}{  \Gamma (p/2) } \sum_{s\neq 0}\left[   \varphi_{p/2-1}( s)
+     \varepsilon ^{(p-D)/2}
   \frac{\pi^{D/2}}{v}
\varphi_{(D-p)/2-1}\left(s
\right)
                             \right] .
 \label{@6}\end{eqnarray}
We have omitted
the $s=0$~-term of the last sum
 since it vanishes for
 $p>D$.
The limit in the brackets
vanishes due to
the
 expansion
(\ref{@xppa}).

Using relation
(\ref{@RELA})
we can rewrite
the Fourier-transformed expression
(\ref{@VPOT})
as
\comment{
\begin{equation}
V_{ij}({\bf k})\equiv
\sum_{\sbf x\neq {\sbf 0}}[ 1-\cos( {\bf k}{\bf x})] \partial _i\partial _j \Phi({\bf x}).
\label{@VPOT2}\end{equation}
Alternatively to the 
Fourier representation (\ref{@3}), we can 
}%
\comment{transform of the potential
$\tilde \Phi({\bf k})$ and
 rewrite(\ref{@3}) as}%
a sum over reciprocal lattice vectors
\begin{equation}
V_{ij}({\bf k})\equiv
\sum_{\sbf c\neq {\sbf 0}}[
({\bf c}+{\bf k})_i
({\bf c}+{\bf k})_j \tilde\Phi(  {\bf c}+{\bf k})
-c_ic_j   \tilde\Phi(  {\bf c})
     ] .
\label{@}\end{equation}
We are now able to
calculate
the effect of the
full lattice sum, splitting
 the
potential in the sum
(\ref{@VPOT})
as in (\ref{@6})
into a small-
and a large-$t$ part
$\Phi^{\sbf x}(r)$ and
$\Phi^{\sbf c}(r)$, so that
\begin{eqnarray}
V_{ij}({\bf k})=
V^{\sbf x}_{ij}({\bf k})+
V^{\sbf c}_{ij}({\bf k})\,,
\label{@}\end{eqnarray}
where
\begin{eqnarray}
V_{ij}^{\sbf x}({\bf k})& =        &
\frac{ \varepsilon ^{p/2}}{ \Gamma (p/2)}
\sum_{\sbf x\neq {\sbf 0}}[ 1-\cos( {\bf k}{\bf x})] \partial _i\partial _j
\varphi_{p/2-1}( \varepsilon {\bf x}^2),\\
V_{ij}^{\sbf x}({\bf k})& =        &
\frac{ \pi^{D/2}
\varepsilon ^{p/2}}{ \Gamma (p/2)v
\varepsilon ^{D/2}}
\sum_{\sbf c}\left[
({\bf c}+{\bf k})_i
({\bf c}+{\bf k})_j
\varphi_{(D-p)/2-1}
\left(
\frac{({\bf k}+{\bf c})^2}{4 \varepsilon }
\right)
-c_ic_j
\varphi_{(D-p)/2-1}
\left(
\frac{{\bf c}^2}{4 \varepsilon }
\right)\right] .
\label{@}\end{eqnarray}
Expanding the sums
up to
powers  $k^4$ and using the property
$\varphi_n'(z)=
-\varphi_{n+1}(z)$,
these become, assuming $(D-p)/2-1<0$,
\begin{eqnarray}
V_{ij}^{\sbf x}({\bf k})& =        &
\frac{ \varepsilon ^{p/2}}{ \Gamma (p/2)}
\sum_{\sbf x\neq {\sbf 0}}\left[
\sfrac{1}{2}
({\bf k}{\bf x})^2
-
\sfrac{1}{24}
({\bf k}{\bf x})^4
		  \right]
\left[4 \varepsilon ^2x_ix_j
\varphi_{p/2+1}
( \varepsilon {\bf x}^2)
 -2 \varepsilon   \delta _{ij}
 \varphi_{p/2}
( \varepsilon {\bf x}^2)
 \right]
,\label{@43}\\
V_{ij}^{\sbf c}({\bf k})
& =        &
V_{ij}^{\sbf c={\sbf 0}}({\bf k})
+
\frac{ \pi^{D/2}\varepsilon ^{p/2}}{ \Gamma (p/2)v
\varepsilon ^{D/2}}
\sum_{\sbf c \neq {\sbf 0}}  \left\{
 ({\bf c}+{\bf k})_i
({\bf c}+{\bf k})_j
\left[
\varphi_{(D-p)/2-1}\left(\frac{{\bf c}^2}{4 \varepsilon } \right)
\right.  \right.\nonumber \\
&&
\left. \!\!\!\!\!\!\!\! \!\!
\left.\!\!\!\!\!\!\!\!\!\!\!\!\!
-\left(\frac{2{\bf c}{\bf k}+{\bf k}^2}{4 \varepsilon }
\right)\varphi_{(D-p)/2}\left(\frac{{\bf c}^2}{4 \varepsilon }\right)
 +\dots
+
\frac{(-1)^4}{4!}\left(\frac{2{\bf c}{\bf k}+{\bf k^2}}{4 \varepsilon }\right)
^4\varphi_{(D-p)/2+3}\left(\frac{{\bf c}^2}{4 \varepsilon }\right)\right]
-c_ic_j
\varphi_{(D-p)/2-1}\left(\frac{{\bf c}^2}{4 \varepsilon }\right)\right\},
\label{@235}\end{eqnarray}
where
$V_{ij}^{\sbf c={\sbf 0}}({\bf k})$ is the purely longitudinal
term
\begin{eqnarray}
V_{ij}^{\sbf c={\sbf 0}}({\bf k})
& =        &
\frac{ \pi^{D/2} \Gamma ((D-p)/2)}
{ \Gamma (p/2)v}{k_ik_j}
\left( \frac{2}{k}\right)^{D-p}
-\frac{2\pi^{p/2}v^{(D-p)/2-1}}
{(D-p) \Gamma (p/2)}
\left[ 1-\frac{D-p}{4(2+D-p)}k^2v+\dots\right]  .
\label{@135}\end{eqnarray}
The higher Misra functions can be reduced to the lower ones by the iteration formula
\begin{equation}
\varphi_{n+1}(z)
=\frac{1}{z}\left[ (n+1)\varphi_{n}(z)+e^{-z}\right] .
\label{@}\end{equation}
We now go to $D=2$ and $p=3$,
and find from
the nearest neighbors
in (\ref{@43})
the sums
\comment{
\begin{eqnarray}
   V_{ij}^{{\sbf x}\,(2)}({\bf k})&=& \frac{3}{2 \sqrt{v} }\left[
 \left(
 \delta _{ij}+2\hat k_i \hat k_j\right)
s^2\varphi_{5/2}(s)-\delta _{ij}\,2
s\varphi_{3/2}(s)\right] ,\\
   V_{ij}^{{\sbf x}\,(4)}({\bf k})&=&
\frac{1}{ 16\sqrt{v} }
\left[
\left(
 \delta _{ij}+4\hat k_i \hat k_j\right)
s^3\varphi_{5/2}(s)-\delta _{ij}\,3
s\varphi_{3/2}(s)\right] ,
\label{@}
\end{eqnarray}
}
\begin{eqnarray}    \!\!\!\!\!\!\!\!\!\!\!\!\!\!\!
   V_{T}^{{\sbf x}\,(2)}({\bf k})\!=\!\sum_{s\neq 0}\frac{\pi s}{v^{3/2}}
\left[3 s\varphi_{5/2}(s)\!-\!6\varphi_{3/2}(s)\right] \approx\frac{6.55}{v^{3/2}},~~
   V_{T}^{{\sbf x}\,(4)}({\bf k})\!=\!\sum_{s\neq 0}\frac{s^2}{16 v^{1/2}}
\left[-s\varphi_{5/2}(s)+6\varphi_{3/2}(s)\right]
 \approx\frac{0.054}{v^{1/2}} ,
\label{@}
\end{eqnarray}
\begin{eqnarray}   \!\!\!\!
   V_{L}^{{\sbf x}\,(2)}({\bf k})\!=\!\sum_{s\neq 0}\frac{\pi s}{v^{3/2}}
\left[9 s\varphi_{5/2}(s)\!-\!6\varphi_{3/2}(s)\right]
\approx\frac{28.47}{v^{3/2}},   ~~
   V_{L}^{{\sbf x}\,(4)}({\bf k})\!=\!\sum_{s\neq 0}\frac{s^2}{16 v^{1/2}}
\left[-{11}s\varphi_{5/2}(s)+6\varphi_{3/2}(s)\right]
\approx-\frac{2.58}{v^{1/2}} .
\label{@}
\end{eqnarray}
If we include all neighbors, the result changes only
little to
\begin{eqnarray}    \!\!\!\!\!\!\!\!\!\!\!\!\!\!\!
   V_{T}^{{\sbf x}\,(2)}({\bf k})
 \approx\frac{7.04}{v^{3/2}},~~
   V_{T}^{{\sbf x}\,(4)}({\bf k})
 \approx\frac{0.080}{v^{1/2}} ,
\label{@}
\end{eqnarray}
\begin{eqnarray}   \!\!\!\!
   V_{L}^{{\sbf x}\,(2)}({\bf k})
\approx\frac{32.53}{v^{3/2}},   ~~
   V_{L}^{{\sbf x}\,(4)}({\bf k})
\approx-\frac{2.59}{v^{1/2}} .
\label{@}
\end{eqnarray}
The
sum over the reciprocal lattice vectors ${\bf c}$-vectors
in (\ref{@235})
has a
purely longitudinal contribution from ${\bf c}=0$:
\begin{eqnarray}
   V_{ij}^{{\sbf c}\,(2)}({\bf k})&=& \frac{k_ik_j}{k^2}
\frac{4 \pi }{ {v}^{3/2} }k^2\left(1-\frac{k \sqrt{v} }{2}+\frac{k^2v}{4\pi}+\dots\right).
 .
\label{@}\end{eqnarray}
Of the remaining sum we include only
the six smallest
${\bf c}$ vectors.
We further approximate this sum
isotropically
by replacing it by $6$ times the angular average
$ \langle \dots\rangle_\phi\equiv
(2\pi)^{-1}\int_0^{2\pi} d\phi \dots$,
where ${\bf c}=(c\cos\phi,\,c\sin\phi)$.
If we define
the subtracted quantities
\begin{eqnarray}
 \gamma _n\equiv
6\langle ({\bf c}+{\bf k})_1({\bf c}+{\bf k})_1 \left(2{\bf c}{\bf k}+{\bf k}^2\right)^n\rangle_\phi
-({\bf k}={\bf 0}),
\label{@}\end{eqnarray}
we obtain
\begin{eqnarray}
 \gamma _0=6 k_1^2,~~~
 \gamma _1=3c^2k^2+(12c^2+6k^2)k_1^2,~~~
 \gamma _2=3c^4k^2+3c^2k^4+(6c^4+36c^2k^2)k_1^2,\\~~~
 \!\!\!\!\!\!\!\!\!\!\!
 \!\!\!\!\!\!\!\!\!\!\!
 \!\!\!\!\!\!\!\!\!\!\!
\gamma _3=9c^4k^4+54c^4k^2k_1^2,~~~~~~~~ \hspace{-1pt}
 ~~~\gamma _4=6c^6k^4+24c^6k^2k_1^2.
~~~~~~~
~~~~~~~~~~~~~~
\label{@}\end{eqnarray}
Note that only $ \gamma _4$ is affected by the isotropic approximation.
The others are the same as in the previous sum over 
only the nearest neighbors.
With these $ \gamma _n$'s
we find from
(\ref{@235})
\begin{eqnarray}
V_{ij}^{\sbf k\neq {\sbf 0}}({\bf k})=
\frac{ \pi^{D/2}\varepsilon ^{p/2}}{ \Gamma (p/2)v
\varepsilon ^{D/2}}
\sum_{q=0}^4
\frac{(-1)^q \gamma _q }{q!(4 \varepsilon) ^p}
\varphi_{(D-p)/2-1+q}(s).
\label{@}\end{eqnarray}
Summing only over the six
nearest neighbors
these become
\begin{eqnarray}
V_T^{{\sbf c}\neq{\sbf 0}} ({\bf k})&=&\frac{2\pi}{v^{3/2} }
k^2\left\{
 \frac{3}{2}\left[
-s\varphi_{-1/2}(s)
+s^2\varphi_{1/2}(s)
\right] +\frac{k^2v}{\pi}
\frac{1}{32}\left[
6s\varphi_{1/2}(s)
-12s^2\varphi_{3/2}(s)
+s^3\varphi_{5/2}(s)
 \right] \right\},                 \\
V_L^{{\sbf c}\neq{\sbf 0}} ({\bf k})&=&\frac{2\pi}{v^{3/2} }
k^2\left\{
 \frac{1}{2}\left[
12 \varphi_{-3/2}(s)
-30s
 \varphi_{-1/2}(s)
+9 s^2\varphi_{1/2}(s)
\right]\right.\nonumber \\&&\left.~~~~~~ +\frac{k^2v}{\pi}
\frac{1}{32}\left[
-48\varphi_{-1/2}(s)
+156\varphi_{1/2}(s)
-84s^2\varphi_{3/2}(s)
+5s^3\varphi_{5/2}(s)
 \right] \right\}.
\label{@}\end{eqnarray}
Inserting $s=2\pi/ \sqrt{3}$
we obtain
\begin{eqnarray}    \!\!\!\!\!\!\!\!\!\!\!\!\!\!
 V_{T}^{{\sbf c}\neq{\sbf 0}}{}^{(2)} =\frac{0.571}{v^{3/2} },~~~~~~
 V_{T}^{{\sbf c}\neq{\sbf 0}}{}^{(4)} =-\frac{0.040}{v^{1/2} },\\\!\!\!\!\!\!\!\!\!\!
 V_{L}^{{\sbf c}\neq{\sbf 0}}{}^{(2)} =\frac{1.019}{v^{3/2} },~~~~~~
 V_{L}^{{\sbf c}\neq{\sbf 0}}{}^{(4)} =-\frac{0.250}{v^{1/2} }.
\label{@}
\end{eqnarray}
Extending the
sum to the entire reciprocal lattice,
these change to
\begin{eqnarray}    \!\!\!\!\!\!\!\!\!\!\!\!\!\!\!
 V_{T}^{{\sbf c}\neq{\sbf 0}}{}^{(2)} =\frac{0.537}{v^{3/2} },~~~~~~
 V_{T}^{{\sbf c}\neq{\sbf 0}}{}^{(4)} =-\frac{0.044}{v^{1/2} },\\\!\!\!\!\!\!\!\!\!\!
 V_{L}^{{\sbf c}\neq{\sbf 0}}{}^{(2)} =\frac{0.831}{v^{3/2} },~~~~~~
 V_{L}^{{\sbf c}\neq{\sbf 0}}{}^{(4)} =-\frac{0.258}{v^{1/2} }.
\label{@}
\end{eqnarray}
Hence we find:
\begin{eqnarray}
						   \ell^2
=0.0041,
\label{@}\end{eqnarray}
which is very small, thus confirming that
the melting
transition will be of first order.

Recently, several different criteria 
for judging the type of melting transitions have been discussed 
in Ref.~\cite{KEIM}
in connection with 
the possibility of studying the melting process 
in two-dimensional suspensions of small colloid spheres \cite{ZAHN}.
Some of them have dipole moments
and shown a hexatic phase. Since the interaction forces in these models
are more complicated   
than the 
pure dipole forces treated here, the order of the transition does not have to 
follow our criterion. 
It will be interesting to understand
the relation between the criteria in \cite{KEIM} 
and the 
simple stiffness 
criterion in \cite{SUCC}.



\begin{thebibliography}{99}



\bibitem{HN}
D.R. Nelson, Phys. Rev. B {\bf 18} (1979) 2318;
D.R. Nelson and B.I. Halperin, Phys. Rev. B {\bf 19} (1979)
2457.
\bibitem{KO}
J.M. Kosterlitz and D.J. Thouless, J. Phys. C {\bf 6} (1973) 1181;
Prog. Low Temp. Phys. B {\bf 7} (1978) 371;
J.M. Kosterlitz, J. Phys. C {\bf 7} (1974) 1046;
\bibitem{SUCC}
 H. Kleinert,
    {\em Lattice Defect Model with Two Successive Melting Transitions\/},
    Phys.~Lett.~A {\bf 130}, 443 (1988)\\
({\tt  http://klnrt.de/174}).
\bibitem{SUCC2}
 H. Kleinert,
    {\em Test of a New Melting Criterion. Angular Stiffness
and Order of 2D Melting in Lennard-Jones and Wigner Lattices\/},
    Phys.~Lett.~A {\bf 130}, 443 (1988)\\
({\tt  http://klnrt.de/183}).
\bibitem{GFCM}
H. Kleinert,
     {\em Gauge Fields in Condensed Matter\/},
     Vol.~II \,\,  Stresses and Defects,
     World Scientific, Singapore 1989 
({\tt http://klnrt.de/b2})


\bibitem{JK}
 W. Janke and H. Kleinert,
    {\em From First-Order to Two Continuous Melting Transitions --
    Monte Carlo
    Study of New $2D$ Lattice Defect Model\/},
    Phys.~Rev.~Lett.~{\bf 61}(20), 2344 (1988)\\
({\tt http://klnrt.de/179})



\bibitem{KEIM}
 P. Dillmann, G. Maret, and P. Keim,
{\it  Comparison  of 2D melting criteria in a
colloidal system},
J. Phys.: Condens. Matter {\bf 24}, 464118
(2012).  



\bibitem{ZAHN}
K. Zahn, R. Lenke, and G. Maret,
Phys. Rev. Lett. {\bf 82}, 2721 (1999).













\end{thebibliography}
\end{document}